# Zero-Class Poisson for Rare-Event Studies

Thomas M. Semkow

Wadsworth Center, New York State Department of Health, Albany, NY 12201, USA

Contact: thomas.semkow@health.ny.gov

**Abstract**

We developed a statistical theory of zero-count-detector (ZCD), which is defined as a zero-class Poisson under conditions outlined in the paper. ZCD is often encountered in the studies of rare events in physics, health physics, and many other fields where counting of events occurs. We found no acceptable solution to ZCD in classical statistics and affirmed the need for the Bayesian statistics. Several uniform and reference priors were studied and we derived Bayesian posteriors, point estimates, and upper limits. It was showed that the maximum-entropy prior, containing the most information, resulted in the smallest bias and the lowest risk, making it the most admissible and acceptable among the priors studied. We also investigated application of zero-inflated Poisson and Negative-binomial distributions to ZCD. It was showed using Bayesian marginalization that, under limited information, these distributions reduce to the Poisson distribution.

## 1. INTRODUCTION

The interest in zero-count detector (ZCD) originates from the studies of rare events. Such studies are often encountered in physics, health physics, and across many fields. The result of searches for rare events may as well be null, however, we are interested in assigning an uncertainty to it. The question is what can we conclude about the underlying process when the result of measurement is null. For a detector to be considered in this category, it has to be reasonably sensitive to the stimulus, however, not all zero results are of interest here. This depends on how the background is measured. The cases satisfying ZCD are listed in Table 1. The net signal is zero in all cases; however, the background has to be either 0, not measured, or rejected on event basis. Specifically, the cases when gross signal and background are measured separately and one of them is zero, or both are equal and non-zero, can be handled by standard statistics and are of no interest here. A more general ZCD can be considered when $n$ zeros are observed in $n$ consecutive measurements.

**Table 1.** Cases satisfying zero-count detector.

| Gross measurement result | Background measurement type | Background result | Net signal |
|---|---|---|---|
| 0 | Measured separately | 0 | 0 |
| 0 | Not measured | Not applicable | 0 |
| 0 | Measured with gross | Rejected by event | 0 |

The goal of analysis of ZCD is to set up an upper limit on a rate of underlying physical process, which may or may not be null. This requires an a-priori statistical model. The most suitable model for this purpose has been the Poisson distribution and its modifications, which allow for zero-class. It has been used extensively in radiation counting, where the number of counts can be low or zero. The Poisson model is not limited to radiation counting, and has been studied in rare-event physics [1], health physics [2], as well applied to zero counts in ecology [3], traffic accidents [4], biology [5], medicine [6], and psychology [7], among others. An alternative model for zero-counts has been studied in optics using Bose-Einstein statistics [8].

## 1.1. The Poisson distribution

We review the properties of the Poisson distribution for subsequent elaboration in this work. The Poisson distribution can be derived from the stochastic Markov process, when there exist a large set of objects $N \gg 1$ and a very small probability $p \ll 1$ of a single successful event. An extensively referenced review has been provided elsewhere [9]. This model is well satisfied by counting of ionizing radiation emitted by a long-lived radionuclide. Then, $N$ is a number of radioactive atoms and $p$ is a probability of a count,

$$p = \lambda t \varepsilon \,, \tag{1}$$

where $\lambda$ is a radioactive decay constant, $t$ is measurement time, and $\varepsilon$ is detection efficiency. The distribution of counts $x$ is given by the Poisson distribution [10]:

$$P(x|\theta) = \theta^x e^{-\theta}/x! \,, \quad x \in [0,1,\dots] \,. \tag{2}$$

where $\theta$ is the parameter given by:

$$\theta = Np \,. \tag{3}$$

The mean counts, its variance, and the dispersion coefficient of counts are given by, respectively:

$$\mu(x) = \theta \,, \tag{4a}$$
$$\mu_2(x) = \theta \,, \tag{4b}$$
$$\delta_x = \mu_2(x)/\mu(x) = 1 \,. \tag{4c}$$

One of the most accurate measurements of the Poisson distribution were reported in gamma radiation counting of $^{137}$Cs ($T_{1/2} = 30.08$ y) source using a NaI detector [11]. A total of 3,109,922 events were collected in 3 h. The events were binned into 10-ms time intervals, yielding the mean counts of 2.8787 and the dispersion coefficient of 1.0017, indicating a nearly perfect Poisson distribution.

In the measurements of physical quantities, we are interested in a steady-state rate $\rho$ of the underlying physical process. For this purpose, the Poisson parameter $\theta$ can be rescaled as follows:

$$\theta = (N\lambda\varepsilon)t = \rho t \,. \tag{5}$$

The variance of the rate is simply given by:

$$\mu_2(\rho) = \rho/t \,. \tag{6}$$

Throughout this work we alternate between the dimensionless count parameter $\theta$ and the rate $\rho$, which has dimension of inverse time.

## 1.2. Deviations from the Poisson distribution

The deviations from the Poisson distribution can be of empirical or fundamental nature and are described in terms of overdispersion, underdispersion, and zero-modified Poisson.

Overdispersion occurs when another fluctuating process overlaps the Markov Poisson process. It can occur quite frequently in actual measurements caused by empirical factors. It has been showed that the dispersion coefficient could be expressed as

$$\delta_x = 1 + \mu_2(x)v^2 > 1 , \tag{7a}$$

where

$$v = (\sigma/\mu)_{\text{ef}} \tag{7b}$$

is the variation coefficient of the excess-fluctuating process [12]. Overdispersion has been studied in electronic pulse pileup [13]. Another reason for overdispersion can be sequential decay [14]. The excess fluctuations can also be of fundamental origins in physics [15]. The Negative-binomial distribution has been successfully used as a model for overdispersed statistics.

In nuclear counting, underdispersion occurs when the rate of the process changes in time owing to radionuclide half-life being comparable with the measurement time. Then the statistics of counts is described by the Binomial distribution. The probability of a count is no longer much less than 1, and the dispersion coefficient $\delta_x = 1 - p < 1$ [9,16]. Another source of underdispersion is disturbance of the Markov process by the dead-time effects [17]. Not fully explained experimental factors caused underdispersion in beta counting [18]. Yet a sub-Poissonian (underdispersed) statistics has been explained by the quantum effects in physics [19].

In practical measurements, one can seldom obtain a perfect agreement with the Poisson distribution described in Ref. [11]. A proper measure of fluctuations is the dispersion coefficient of counts, determined experimentally. As an example, the dispersion coefficients of background counts were measured on a liquid scintillation counter [20]. For every set, at least 100 measurements were made, of 100 min duration each, resulting in the dispersion coefficients ranging from 0.78 to 1.24. If the process studied can be nevertheless treated as a steady state, then its variance can be modified from Eq. (6) as follows [21]:

$$\mu_2(\rho) = \delta_x \rho/t . \tag{8}$$

Another deviation from the Poisson distribution of interest to ZCD is zero-modified Poisson, which can be either zero-inflated or zero-deflated [10]. The former is more common and has been used extensively to study zero-class [3-7].

The question is: can the deviations from the Poisson distribution have any effect on ZCD? The Binomial distribution is of less interest to ZCD in nuclear counting because the rate of rare events may not be rapidly varying. The potential applicability of Negative-binomial and zero-inflated Poisson distributions to zero-class of rare events will be described in Section 4.

## 1.3. Approaches to zero-count detector

The goal of statistical analysis of ZCD is to set up an upper limit on the rate of the underlying physical phenomenon with the limited information available, consisting only of: zero counts detected in a given measurement time as well as assumption of the Poisson process. The following approaches will used for this purpose:

- 1-count upper limit
- classical statistical analysis
- Bayesian analysis.

The 1-count upper limit is described in the Section 5. The classical statistical approach is outlined Section 2, where we describe the likelihood function and sufficient statistics in Section 2.1, and unbiased statistics in Section 2.2. Additionally, we perform statistical estimations using the methods of maximum likelihood and simple probability (Section 2.3). In Section 3, we describe the principles of Bayesian statistics (Section 3.1) and discuss the concept of prior probability in Section 3.2 in detail. Subsequently, we derive Bayesian posterior and estimates in Section 3.3, followed by discussing Bayesian acceptance in terms of bias, risk, and admissibility in Section 3.4. Section 5 is devoted to the discussion and interpretation of the above concepts and their applications to ZCD. Since many of the original derivations are scattered in the vast statistical literature, simplified derivations are often provided for the purpose of this work to justify the concepts investigated.

## 2. CLASSICAL STATISTICAL ANALYSIS

### 2.1. Poisson likelihood function and sufficient statistics

We begin with a formal classical estimation from the Poisson distribution using maximum likelihood (ML) method. Let us assume that $n$ samplings (observations or measurements) were made from the Poisson distribution given by Eq. (2) resulting in a vector of counts $\boldsymbol{x} = \{x_1, \ldots, x_n\}$, regardless of whether the individual counts $x_i$ were zero or positive. The sum of all counts is given by:

$$S = \sum_{i=1}^{n} x_i \, , \tag{9}$$

and the sample mean counts by:

$$\bar{x} = S/n \, . \tag{10}$$

The likelihood function is proportional to the product of individual distributions from Eq. (2) [22]:

$$L(\boldsymbol{x}|\theta) \propto \frac{\theta^S e^{-n\theta}}{\prod_{i=1}^{n} x_i!} = g(S|\theta)h(\boldsymbol{x}) \propto (\rho t)^S e^{-n\rho t}. \tag{11}$$

In classical statistics, the likelihood function is often abbreviated as $L(\theta|\boldsymbol{x})$, reflecting the fact that $\theta$ is determined by ML given the data $\boldsymbol{x}$. However, Bayesian statisticians often use $L(\boldsymbol{x}|\theta)$, because it is consistent with the inversion in the Bayes theorem described later. While the likelihood in Eq. (2) is normalized, the generalized likelihood is Eq. (11) is not normalized. Since $L$ can be factored in two functions: $g$ dependent on $S$ and $\theta$ but independent on $x_i$, and $h$ dependent on $x_i$ but independent of $\theta$, $S$ is called sufficient statistics [23]. For only a single measurement made, sufficient statistics does not enter the picture.

By differentiating of $\ln L$ from Eq. (11) with respect to $\theta$, setting it to zero, and using Eq. (10), one obtains the ML estimate of $\theta$ [10]:

$$\hat{\theta} \equiv \hat{\mu}(x) = \bar{x} \,. \tag{12}$$

## 2.2. Unbiased statistics

An estimator $\beta$ of parameter $\theta$ is called unbiased if its expectation value $E$ is equal to [23]:

$$E[\beta] = \theta \,. \tag{13}$$

Since the first uncorrected moment of the Poisson distribution is given by [10]:

$$E[x] = \sum_{x=0}^{\infty} x P(x|\theta) = \theta \,, \tag{14}$$

therefore

$$E[\bar{x}] = \theta \,. \tag{15}$$

Consequently, the ML estimator $\hat{\theta}$ in Eq. (12) is unbiased. The unbiased ML estimator of the rate is given by:

$$\hat{\rho} = S/(nt) \,. \tag{16}$$

The second corrected moment (the variance) of the Poisson distribution is given by [10]:

$$E[(x-\theta)^2] = \sum_{x=0}^{\infty} (x-\theta)^2 P(x|\theta) = \theta \,. \tag{17}$$

Therefore, the ML unbiased estimators for the variances are given by:

$$\hat{\mu}_2(x) = S/n \,, \qquad \text{of the counts} \tag{18a}$$
$$\hat{\mu}_2(\bar{x}) = S/n^2 \,, \qquad \text{of the mean counts} \tag{18b}$$
$$\hat{\mu}_2(\rho) = S/(nt)^2 \,. \qquad \text{of the rate} \tag{18c}$$

The physical meaning of sufficient statistics $S$ can be explained using Eq. (18c). It is seen that the variance of the rate is the same whether performing $n$ measurements of duration $t$ each and summing up the individual counts to $S$, or performing a single measurement of duration $nt$ with the total counts $S$. The sufficient statistic also pertains to the estimate of rate in Eq. (16).

If we did not consider the properties of the Poisson distribution, and evaluated a sample variance of counts, $(1/n)\sum(x-\bar{x})^2$, it would be a biased estimator. Replacing $n$ with $n-1$ in the denominator makes the sample variance unbiased [23,24].

## 2.3. Simple probability

A classical approach to ZCD considers probability of zero-class. Using Eq. (2), the probability of zero-class is given by:

$$P(0|\theta) = e^{-\theta}. \tag{19}$$

For the experimental example [11] given above, $P(0|2.8787) \cong 5.621 \times 10^{-2}$.

Johnson [25] studied expectation of zero-class using Eq. (19). His method requires two or more measurements, at least one nonzero. Therefore, this method does not apply to ZCD studied here, where we assume that only zero counts were detected.

However, one can assume *ad-hoc* that Eq. (19) represents a normalized probability of $\theta$ [26, 27]. More generally, if $n$ measurements resulted in zero counts each, the normalized probability of zero class would be given by:

$$P(\mathbf{0}|\theta) = ne^{-n\theta}. \tag{20}$$

Using Eqs. (5) and (20), the mean and variance of $\rho t$ are given by:

$$\mu(\rho t) = 1/n \, , \tag{21a}$$
$$\mu_2(\rho t) = 1/n^2 \, . \tag{21b}$$

One can establish an upper limit on $\theta = \rho t$, given a confidence level (CL) of $\rho t$ being below it in a random experiment and a significance of $\alpha = 1 - \mathrm{CL}$ that $\rho t$ exceeds it in a random experiment. The upper limit is obtained by integration of Eq. (20):

$$\alpha = \int_{U_\theta}^{\infty} P(\mathbf{0}|\theta)d\theta \, ,$$

resulting in:

$$U_\theta = \frac{1}{n}\ln\left(\frac{1}{\alpha}\right) , \tag{22a}$$
$$U_\rho = U_\theta/t \, . \tag{22b}$$

These results will be interpreted in the Section 5.

## 3. BAYESIAN ANALYSIS

### 3.1. The principles of Bayesian statistics

There are several established texts describing Bayesian statistics [28-32] with extensive lists of references. The purpose of this Section is to summarize key concepts which will be used in this paper. A 1-dimensional version of the Bayes theorem is given below:

$$P(\theta|\mathbf{x}) = \frac{P(\mathbf{x}|\theta)P(\theta)}{\int P(\mathbf{x}|\theta)P(\theta)d\theta} \, . \tag{23}$$

In Eq. (23), $P(\mathbf{x}|\theta)$ is the likelihood function, such as given by Eq. (2) or (11), and $P(\theta)$ is a prior probability described in the Section 3.2. The denominator is an integral over the numerator and it is called a Bayesian evidence. This integral ensures that the posterior $P(\theta|\mathbf{x})$ is a normalized probability. Being continuous in parameter $\theta$, the posterior is a probability density function (pdf).

The Bayes theorem in Eq. (23) inverts the likelihood function. The key advantage of the Bayesian statistics is that the posterior is now a function of the parameter sought given the data, from which one can infer its moments, credibility intervals, and upper limits. This is only when the prior probability is proper, so that the normalization integral and moment integrals converge.

Since the original Bayes theorem [33] and its mathematical reformulation by Laplace [34], one of the strongest modern developers of Bayesian statistics was Jeffreys [35-37]. In the area of statistical physics, significant advance of Bayesian statistics has been attributed to Jaynes [37,38]. In this paper, we are particularly interested in the application of Bayesian statistics to physics [39,40], as well as health physics and radiation measurements [22,41,42].

## 3.2. Prior probability

The main difference between the classical and Bayesian statistics is in the former assigning the probability to frequency, whereas the latter to knowledge (or information). This knowledge is built into the prior probability, and is the most debated aspect of the Bayesian statistics. In this Section, we discuss various prior probabilities often providing simplified derivations.

The simplest prior is a uniform prior. Another class of priors are hierarchical and empirical priors, where previous knowledge from experimental observations is incorporated into the prior of some functional form for subsequent inference. For mathematical convenience, often a conjugate form of the prior is selected which results in the posterior having the same functional form as the prior. In case of studying rare events where zero is observed, such prior information is usually not available except, possibly, another zero.

Therefore, for the purpose of ZCD one needs to consider a class of reference (or noninformative or indifferent) priors. They assume complete prior ignorance of the parameter(s) sought. They are not valid if any additional prior information about the data is known. However, the noninformative priors are based on some generally recognized principle, law, or criterion. They are also referred to as objective priors in a sense of the principle of indifference. One form of this principle is that the two observers having the same knowledge have to assign the same probability to it [43]. Therefore, if a recognized principle was provided to independent observers, it would result in these observers arriving at the same prior, thus making it more objective.

*Uniform prior*

The simplest prior is obtained when there is no a priori knowledge or principle about parameter $\theta$. This leads to the prior

$$P(\theta) = P(\rho) = \text{const.}, \tag{24}$$

which is called the uniform or Bayes/Laplace (BL) prior. By using the Bayes theorem, Eq. (23), with the likelihood function from Eq. (11) and the uniform prior, Eq. (24), we obtain the posterior density [44]:

$$P(\theta|\boldsymbol{x}) = \frac{n(n\theta)^S e^{-n\theta}}{\Gamma(S+1)}, \tag{25}$$

where $\Gamma$ is the Gamma function [45]. Equation (25) can be expressed as a probability density of the rate:

$$P(\rho|\boldsymbol{x}) = \frac{nt(n\rho t)^S e^{-n\rho t}}{\Gamma(S+1)}. \tag{26}$$

It should be noted that the uniform prior is not invariant under transformation. Namely, by taking $\rho' = \rho/q$, and $P(\rho) = P(\rho') = c$, where $q$ and $c$ are arbitrary constants, one obtains:

$$P(\rho)d\rho = cqd\rho' = qP(\rho')d\rho' \neq P(\rho')d\rho' \,.$$

However, the posterior remains invariant.

*Transformation invariant priors*

The transformation priors are based on the principle of invariance under transformation. The original derivation was given by Jeffreys for the prior of the $\sigma$ parameter of the normal (Gaussian) distribution [46] resulting in the prior $\propto d\sigma/\sigma$. In his book on probability [35,36], Jeffreys showed that this condition is equivalent to $\log \sigma \sim$const. Still another derivation assumed the probability of 1/3 that a third random observation would lie between the first and the second [47]. Jeffreys distinguished location parameters (such as the normal $\mu$) from the scale parameters (such as the normal $\sigma$). According to his theory, the corresponding priors are:

$$P(\mu) \propto \text{const.}, \tag{27a}$$
$$P(\sigma) \propto 1/\sigma. \tag{27b}$$

If a normal likelihood function, $L(\boldsymbol{x}|\mu,\sigma)$ is marginalized with these priors, over $\mu$ and $\sigma$ separately, one obtains the $\chi^2$- and $t$-distributions, respectively [48].

For the Poisson likelihood, Eq. (11), one has only the scale parameter $\rho$. Jaynes [49] provided a rigorous derivation of this prior based on a group transformation. We outline a simplified derivation below. Let us suppose that time $t$ has been rescaled to $t' = qt$, where $q$ is a constant. In both time scales, we have to have Poisson count parameter the same, so $\rho't' = \rho t$, from which $\rho' = \rho/q$. We also require the Bayes posterior $P(\rho|\boldsymbol{x})d\rho$ from Eq. (23) invariant under the time transformation. Since the likelihood function (Eq. (11)) does not change by the transformation, we must have for the prior:

$$P(\rho)d\rho = P'(\rho')d\rho' = P'(\rho')\frac{d\rho}{q},$$

from which:

$$qP(\rho) = P'(\rho') \,.$$

By assuming a trial solution of the form $P(\rho) \sim \rho^d$, where $d$ is an exponent, we obtain from the above equation:

$$qP(\rho) = q\rho^d = q^{d+1}(\rho')^d = P'(\rho') \,,$$

only if $d = -1$. Therefore, the prior for the Poisson rate invariant under the time group transformation is given by

$$P(\rho) \propto 1/\rho \,. \tag{28}$$

Still, another derivation was provided by Jaynes [43] using marginalization.

The prior as an inverse of the scale parameter is called Jeffreys prior in the statistical literature. However, to distinguish it from other priors and to recognize Jaynes [49] derivations using group transformation, we will call it Jeffreys-Jaynes (JJ) prior. Application of the group transformation to the binomial's likelihood $L(\boldsymbol{x}|N,p)$ parameter $p$ results in a prior $\sim p^{-1}(1-p)^{-1}$ [49], in agreement with Haldane [50].

Furthermore, Jeffreys studied a class of prior invariant under a one-to-one parameter transformation [35, 51], arriving at yet another form of Jeffreys prior. An independent derivation was given by Perks [52]. A simplified derivation following Gelman et al. [32] is reproduced below.

Let us consider a one-to one transformation of continuous parameter $\theta$ into $\varphi$, which are connected with some function $h$: $\varphi = h(\theta)$, $\theta = h^{-1}(\varphi)$. One requires that the prior $P$ of these parameters be invariant under the transformation, i.e.,

$$P(\varphi)d\varphi = P(\theta)d\theta \,,$$

from which,

$$P(\varphi) = P(\theta) \left| \frac{d\theta}{d\varphi} \right| . \tag{29}$$

In a modern notation, let us consider Fisher information $J(\varphi)$:

$$J(\varphi) = -E\left[ \frac{d^2 \log P(\boldsymbol{x}|\varphi)}{d\varphi^2} \right] , \tag{30}$$

from which one obtains by substitution:

$$J(\varphi) = -E\left[ \frac{d^2 \log P(\boldsymbol{x}|\theta = h^{-1}(\varphi))}{d\theta^2} \left| \frac{d\theta}{d\varphi} \right|^2 \right] = J(\theta) \left| \frac{d\theta}{d\varphi} \right|^2 . \tag{31}$$

The square root of Eq. (31) yields:

$$J(\varphi)^{1/2} = J(\theta)^{1/2} \left| \frac{d\theta}{d\varphi} \right| . \tag{32}$$

By comparing Eq. (29) with Eq. (32), one can conclude that the prior invariant under the one-to-one transformation is proportional to the square root of the Fisher information, Eq. (30):

$$P(\theta) \propto J(\theta)^{1/2} \,, \tag{33}$$

which is the Jeffreys rule (JR). In his derivation, Jeffreys used an older notation without invoking Fisher information explicitly.

By applying the JR to the normal likelihood $L(\boldsymbol{x}|\mu, \sigma)$ for $\mu$ and $\sigma$ separately, one obtains the priors $P(\mu) \propto$ const. and $P(\sigma) \propto 1/\sigma$ in agreement with the group-transformation priors discussed above (Eq. (27)). Jeffreys gave the reasons that this was an acceptable solution although by no means the only one. Alternatively, when applying the JR to a two-dimensional problem for $(\mu, \sigma)$ simultaneously, one obtains a combined prior $P(\mu, \sigma) \propto 1/\sigma^2$. Robert et al. [36] considered an example of $m$ groups of normal observations with $n$ measurements in each group, having different means $\boldsymbol{\mu} = \{\mu_1, \ldots, \mu_m\}$, and a common standard deviation. The combined multi-dimensional prior is then $P(\boldsymbol{\mu}, \sigma) \propto 1/\sigma^{m+1}$. If $n$ is smaller against $m$, the expectation of Bayes posterior on $\sigma$ significantly deviates from the true standard deviation. Therefore, such multi-dimensional priors are unacceptable.

In this work, we are interested in the Poisson likelihood. When applying JR to the Poisson likelihood $L(\boldsymbol{x}|\rho)$ given by Eq. (11), one obtains from Eq. (30):

$$J(\rho) = -E\left[\frac{d^2 \log L(\boldsymbol{x}|\rho)}{d\rho^2}\right] = \frac{E[S]}{\rho^2} = \frac{n}{\rho},$$

from which the prior is proportional to

$$P(\rho) \propto 1/\rho^{1/2}, \tag{34}$$

by means of Eqs. (33) and (5). This prior differs from the prior $\propto 1/\rho$ by the group transformation, which will have a significant effect on the inference for zero counts. Unfortunately, Eq. (34) does not carry the transformation invariance of the JJ prior (Eq. (28)).

*Maximum entropy prior*

Jaynes [43] reviewed a progression of the maximum entropy (ME) principle in statistical physics from Boltzmann to Maxwell and Gibbs, and to information theory of Shannon. The priors based on the maximum entropy (or information) can also be considered reference priors [53]. The original entropy expression, $W = -\sum p \log p$, was designed for discrete distributions. For continuous distributions, it lacks invariance and needs to be modified to

$$W = -\int p(\rho) \log\left[\frac{p(\rho)}{h(\rho)}\right] d\rho, \tag{35}$$

where $p(\rho)$ is a probability density function and $h(\rho)$ is an invariant measure [49,54-56]. A derivation using Lagrange multipliers is presented below.

We start with the entropy to be maximized in Eq. (36a) and constraint conditions: normalization in Eq. (36b) and a finite mean rate $\rho_0$ in Eq. (36c).

$$W = -\int p(\rho) \log p(\rho) \, d\rho, \tag{36a}$$

$$\int p(\rho) d\rho = 1, \tag{36b}$$

$$\int \rho \, p(\rho) d\rho = \rho_0. \tag{36c}$$

A functional $F$ is defined, using Eqs. (36):

$$F(p) = W - \kappa \cdot 1 - \nu\rho_0 = -\int_0^\infty (p \log p + \kappa p + \nu p\rho) d\rho,$$

where $\kappa$ and $\nu$ are Lagrange multipliers. The variational derivative of $F$, $\delta F$ is equal to

$$\delta F = -\int_0^\infty (\log p + 1 + \kappa + \nu\rho) d\rho \delta p.$$

The derivative $\delta F = 0$ if $\log p + 1 + \kappa + \nu\rho = 0$, from which:

$$p(\rho) = e^{-1-\kappa} e^{-\nu\rho}. \tag{37}$$

By inserting Eq. (37) into Eqs. (36b) and (36c), one obtains:

$$p(\rho) = e^{-\rho/\rho_0}/\rho_0.$$

The best estimate of the mean rate is $\rho_0 = 1/t$ in a measurement lasting $t$. Therefore, the ME prior is given by:

$$p(\rho)d\rho = te^{-\rho t}d\rho \,, \tag{38}$$

and it is invariant under time transformation.

Equation (35) is satisfied for $h(\rho) = e^{-\rho t}$ and $p(\rho) = h(\rho)/\int h(\rho)$, in agreement with Ref. [49]. It is interesting that Fisher [57] used an *ad-hoc* defined prior given by Eq. (38) in his polemics with Jeffreys. If variable and parameter are interchanged in Eq. (38), it represents a time distribution between events in a stochastic Markov process [11].

*Additional priors*

Additional studies on priors are described in the textbooks [28-32]. The selected works on the reference priors are those by Bernardo [53], and Kass and Wasserman [58]. Specific priors of interest are asymptotically locally invariant priors [59], locally impartial priors [60], neutral noninformative priors [61], and convergent noninformative priors [22]. Irony [62] compiled the priors for the discrete distributions.

### 3.3. Bayesian posterior and estimators

A conjugate prior is defined as the one which results in the posterior having the same functional form as the prior. It is well recognized in the statistical literature that Gamma prior is conjugate for the Poisson distribution [62]. We use the following convention for the Gamma distribution prior density for $\rho$, slightly different from that by Johnson et al. [63]:

$$P(\rho|a,b) = \frac{b^a}{\Gamma(a)} \rho^{a-1}e^{-b\rho} \propto \rho^{a-1}e^{-b\rho}, \tag{39}$$

where $a > 0$ and $b > 0$ are parameters (the limits when $a$ or $b$ are zero will be discussed later). The uncorrected $r$-th moment is given by:

$$\mu_r' = \int_0^\infty \rho^r P(\rho|a,b)d\rho = \frac{(a)_r}{b^r} \,, \tag{40}$$

where $(a)_r = a(a+1), \ldots, (a+r-1)$ is an ascending factorial, $(a)_0 = 1$ [10]. The mean and the variance are given by:

$$\mu(\rho) \equiv \mu_1' = a/b \,, \tag{41a}$$
$$\mu_2(\rho) = \mu_2' - \mu^2 = a/b^2 \,, \tag{41b}$$

An upper limit on the rate, $U_\rho$ can be calculated similarly to Eq. (22), by a numerical solution of Eq. (42) below:

$$\mathrm{CL} = 1 - \alpha = \frac{\gamma(a, bU_\rho)}{\Gamma(a)} \,, \tag{42}$$

where $\gamma$ is an incomplete Gamma function [45].

It has been showed that the uniform and reference priors can be expressed as special cases of the term proportional to the Gamma distribution given by Eq. (39) [22]. This is reproduced in Table 2, where the rows correspond to the priors considered: uniform (BL), Eq. (24), as well as reference (JJ, JR, ME), Eqs. (28), (34), and (38), respectively. The prior formulas are given together with the principles satisfied. The values of $a$ and $b$ parameters are also given in Table 2.

**Table 2.** Definition of priors for the Bayesian analysis from the Poisson likelihood.

| Prior | | | Underlying principle | | | | Parameter | |
|---|---|---|---|---|---|---|---|---|
| Category | Symbol | Formula | Unbiased | Invariant | Fisher information | Maximum entropy | $a$ | $b$ |
| Uniform | BL | const. | | | | | 1 | 0 |
| Reference | JJ | $1/\rho$ | yes | yes | | | 0 | 0 |
| | JR | $1/\rho^{1/2}$ | | | yes | | 1/2 | 0 |
| | ME | $e^{-\rho t}$ | | yes | | yes | 1 | $t$ |

Returning to the Bayesian analysis, we calculate the posterior from Eq. (23) expressed as a function of the rate $\rho$, with the Poisson likelihood (Eq. (11)), using Gamma prior (Eq. (39)). The posterior density is, as expected, another Gamma distribution:

$$P(\rho|A,B) = P(\rho|S,n,t;a,b) = \frac{B^A}{\Gamma(A)}\,\rho^{A-1}e^{-B\rho}\,, \tag{43}$$

where $A = S + a > 0$ and $B = nt + b > 0$. The uncorrected $r$-th moment, mean, and variance are generalizations of Eqs. (40) and (41), resulting in the Bayesian point estimators [22,62]:

$$\mu'_r(\rho) = \frac{(S+a)_r}{(nt+b)^r}\,, \tag{44a}$$

$$\mu(\rho) = \frac{S+a}{nt+b}\,, \tag{44b}$$

$$\mu_2(\rho) = \frac{S+a}{(nt+b)^2}\,. \tag{44c}$$

The sufficiency of statistics is evident in Eqs. (44).

The upper limit can be calculated by solving:

$$\mathrm{CL} = 1 - \alpha = \frac{\gamma\big[S+a,(nt+b)U_\rho\big]}{\Gamma(S+a)}\,, \tag{45}$$

for a desired confidence level CL (which is termed a credibility level in Bayesian inference), or Bayesian significance $\alpha$.

### 3.4. Bayesian acceptance

The acceptance of Bayesian estimates, and thus corresponding priors and posteriors, can be studied in terms of bias, risk, and admissibility [24,28,31], in spite of them being classical-statistics concepts.

Let us consider a risk function when using a $\beta$ estimate of the parameter $\theta$. This risk function is usually evaluated for a quadratic loss function, $(\beta - \theta)^2$, integrated or summed-up over the sampling distribution, resulting in [24]:

$$\text{Risk}(\theta_\beta) = (E[\beta] - \theta)^2 + \text{Var}(\beta) \,. \tag{46}$$

The optimal estimator, called an unbiased minimum variance estimator (UMV), is when the first term on the right-hand side in Eq. (46) is zero (corresponding to an unbiased estimator from Eq. (13)), and the second term is minimum.

The unbiased statistics remains one of the most used criteria in evaluating acceptance of the estimates in classical statistics. However, Jaynes [24] provided an example that, when the estimator is unbiased, the second term in Eq. (46) may increase, so the sum of both terms (i.e., the risk) may increase. Therefore, by classical definition, this unbiased estimator would be inadmissible (i.e., would not have the lowest risk [30]). On the contrary, when the estimate is biased, the first term is positive but the second term may decrease, and the overall risk as a sum of both terms actually decreases. Therefore, the unbiased estimator may not provide the lowest risk under all circumstances, and may not always be admissible.

In the following, we calculate the bias and risk of the Bayesian estimate of the mean counts, $\theta_B$, which is converted from Eq. (44b) by setting $t = 1$:

$$\theta_B = \frac{S + a}{n + b} \,. \tag{47}$$

Working in the space of counts enables straightforward calculations of the expectation value of a desired moment of the estimate, $\beta^m$, which is defined for the purpose of risk over the Poisson likelihood as:

$$E[\beta^m] = \sum_{x=0}^{\infty} \beta^m P(x|\theta) \,. \tag{48}$$

The bias is calculated as:

$$\text{Bias}(\theta_B) = E[\theta_B] - \theta = \frac{a - b\theta}{n + b} = \frac{a - bS/n}{n + b} \,, \tag{49}$$

where we substituted $\theta$ by its ML estimate of $S/n$ from Eqs. (10) and (12). For the purpose of the risk, the variance is calculated as:

$$\text{Var}(\theta_B) = E[\theta_B^2] - (E[\theta_B])^2 = \frac{n\theta}{(n + b)^2} \,. \tag{50}$$

By combining Eqs. (46), (49), and (50), the risk is calculated as:

$$\text{Risk}(\theta_B) = \text{Bias}(\theta_B)^2 + \text{Var}(\theta_B) = \frac{S + (a - bS/n)^2}{(n + b)^2} \,. \tag{51}$$

The Bayesian estimate of variance is converted from Eq. (44c) by setting $t = 1$:

$$V_B = \frac{S + a}{(n + b)^2} \,. \tag{52}$$

Repeating the calculations, one obtains for the bias:

$$\text{Bias}(V_B) = \frac{S+a}{(n+b)^2} - \frac{S}{n} \,,\qquad (53)$$

and for the risk

$$\text{Risk}(V_B) = \left[\frac{S+a}{(n+b)^2} - \frac{S}{n}\right]^2 + \frac{S}{(n+b)^4} \,.\qquad (54)$$

## 4. OTHER DISTRIBUTIONS

Two other distributions are investigated below: zero-inflated Poisson (z-Poisson) and Negative binomial (NB) for their potential applications to ZCD. In order to simplify mathematical treatment, we will generally assume $n = t = 1$, which implies $S = x$.

### 4.1. Zero-inflated Poisson

We propose a new formula for the z-Poisson distribution, which offers a simpler mathematical handling of Bayesian estimation than the textbook formula [10]:

$$P_z(x|\theta,\psi) = \begin{cases} P[x=0] = \psi P_0 \\ P[x=j\geq 1] = \dfrac{1-\psi P_0}{1-P_0} P_j \end{cases} ,\qquad (55)$$

where $\psi$ is a parameter ($1 < \psi < 1/P_0$), and $P_j$ is a shorthand notation for the Poisson distribution from Eq. (2), $P(j|\theta)$. The mean and dispersion coefficient are given by, respectively:

$$\mu(x) = \frac{1-\psi P_0}{1-P_0}\,\theta \,,\qquad (56a)$$

$$\delta_x = 1 + \frac{(\psi-1)P_0}{1-P_0}\,\theta \,.\qquad (56b)$$

Therefore, the z-Poisson distribution is overdispersed and it reduces to the Poisson distribution when $\psi = 1$.

We proceed with the Bayesian analysis by assuming ME priors for the $\theta$ and $\psi$ parameters (compare Eq. (38)). The posterior is given by a 2-dimensional version of Eq. (23):

$$P(\theta,\psi|x) = \frac{P_z(x|\theta,\psi)e^{-\theta}e^{-\psi}}{\int_0^\infty d\theta \int_0^\infty d\psi\, P_z(x|\theta,\psi)e^{-\theta}e^{-\psi}} = \begin{cases} 2\psi P_0\, e^{-\theta}e^{-\psi} \,, & x=0 \\ \dfrac{1-\psi P_0}{1-P_0} 2^{j+1} P_j e^{-\theta}e^{-\psi} \,, & x=j\geq 1 \end{cases} .\qquad (57)$$

If $\psi$ parameter is considered unknown, the Bayesian principle allows marginalizing over it (i.e., integrating it out). The result is given below:

$$P(\theta|x) = \int_0^\infty P(\theta,\psi|x)\, d\psi = 2(2\theta)^x e^{-2\theta}/x! \,,\qquad (58)$$

which is the posterior for the Poisson likelihood with ME prior, Eq. (43).

### 4.2. Negative binomial

There are several interpretations of the NB distribution. The textbook version emphasizes interpretation as an inverse of the Binomial distribution [10]. However, the NB distribution can also be considered as a Gamma mixture of the Poisson distribution, where Poisson mean fluctuates according to the Gamma distribution, Eq. (39). This leads to the overdispersed statistics [12,22]. These two versions have different parametrizations in the two parameters. For the purpose of this investigation, we parametrize the NB with $(\theta, a)$ where we label the distribution mean as $\mu \equiv \theta$ to emphasize the relationship with the Poisson distribution.

$$\text{NB}(x|\theta, a) = \frac{\theta^x (a)_x}{x! \, a^x (1 + \theta/a)^{x+a}} \, , \quad x \in [0, 1, \dots] \, . \tag{59}$$

The dispersion coefficient is given by:

$$\delta_x = 1 + \theta/a \, , \tag{60}$$

in agreement with Eq. (7a).

Similar to the z-Poisson, we assume ME priors for the $\theta$ and $a$ parameters, and use hypergeometric-series expansion [10]:

$$\frac{1}{(1 + \theta/a)^{x+a}} = {}_2F_1[x + a, b; b; -\theta/a] = \sum_{k=0}^{\infty} \frac{(x + a)_k (b)_k}{(b)_k} \frac{(-\theta/a)^k}{k!}$$

$$= \sum_{k=0}^{\infty} (x + a)_k \frac{(-\theta/a)^k}{k!} \, , \tag{61}$$

where ${}_2F_1$ is the Gaussian hypergeometric function. The posterior density is then given by:

$$P(\theta, a|x) = \frac{\text{NB}(x|\theta, a) e^{-\theta} e^{-a}}{\int_0^{\infty} d\theta \int_0^{\infty} d\psi \, \text{NB}(x|\theta, a) e^{-\theta} e^{-a}} = \frac{\text{NB}(x|\theta, a) e^{-\theta} e^{-a}}{\sum_{k=0}^{\infty} (x + 1)_k (-1)^k I_{x+k}/k!} \, , \tag{62}$$

where $I_{x+k}$ integral is defined below.

By marginalizing over the unknown parameter $a$, we obtain:

$$P(\theta|x) = \frac{\theta^x e^{-\theta}}{x!} \frac{\sum_{k=0}^{\infty} (-\theta)^k I_{x+k}/k!}{\sum_{k=0}^{\infty} (x + 1)_k (-1)^k I_{x+k}/k!} \, . \tag{63}$$

The relation to the Poisson distribution in Eq. (2) is clearly recognized in Eq. (63). This equation would contain two divergent series: exponential in the numerator and hypergeometric in the denominator, which could be summed up, if not the integrals $I_{x+k}$.

Using:

$$\frac{(a)_{x+k}}{a^{x+k}} = \frac{a(a + 1) \cdots (a + x + k - 1)}{a^{x+k}} = \prod_{y=1}^{x+k-1} \left(1 + \frac{y}{a}\right) = 1 + \sum_{y=1}^{x+k-1} \frac{c_{ky}}{a^y} \, ,$$

where $c_{ky}$ are expansion coefficients, the integral $I_{x+k}$ is given by:

$$I_{x+k} = \int_0^\infty \frac{(a)_{x+k}}{a^{x+k}} e^{-a} da = 1 + K_{x+k} \,, \tag{64a}$$

$$K_{x+k} = \sum_{y=1}^{x+k-1} c_{ky} D_y \,, \tag{64b}$$

$$D_y = \int_0^\infty \frac{e^{-a}}{a^y} da \,. \tag{64c}$$

Note that, $I_0 = I_1 = 1$, however $I_{x+k}$ is generally singular at $a = 0$, owing to the presence of divergent integrals of the $y$-th order $D_y$ defined by Eq. (64c).

In an attempt to remove the singularities, we plug Eq. (64a) into Eq. (63) and sum up the series, which results in in the marginal distribution:

$$P(\theta|x) = \frac{\theta^x e^{-\theta}}{x!} \frac{e^{-\theta}}{1/2^{x+1}} \frac{1 + e^\theta \sum_{k=0}^\infty (-\theta)^k K_{x+k}/k!}{1 + 2^{x+1} \sum_{k=0}^\infty (x+1)_k (-1)^k K_{x+k}/k!} = 2(2\theta)^x e^{-2\theta}/x! \,. \tag{65}$$

Because the divergent terms $K_{x+k}$ are the same in the numerator and denominator, and are normalized by the exponential and hypergeometric functions, respectively, they are expected to cancel out in Eq. (65). We are left, as in Eq. (58), with the posterior of the Poisson likelihood with the ME prior, Eq. (43).

## 5. DISCUSSION

Even if 0 counts were detected, the 1-count upper limit assumes that a single count was detected in the measurement [27]. The 1-count upper limit is not a statistical parameter and does not carry confidence level. When converted to a rate, using measurement time and calibration coefficient, it provides a measure of detection capability for the rate, and it is easy to calculate.

The classical ML estimators for the mean counts (Eq. (12)) and the rate (Eq. (16)) are unbiased and they result in zero when $S = 0$. This in itself would be acceptable since there is no guarantee that the effect under study is positive. Unfortunately, all ML unbiased variance estimators (Eqs. (18)) are also zero when $S = 0$. This is a pathological result. The goal of inference from ZCD is to provide some measure of uncertainty using the Poisson model. Since the estimate of uncertainty on the null results is not available, the ML estimate is not an acceptable solution for ZCD.

Turning now to an *ad-hoc* simple probability method, it is seen from Eqs. (21) that, when observed counts are zero, the estimates of the mean count and variance are non-zero. These estimates are biased, and their values decrease when the number of null measurements $n$ increases, which is intuitively expected. For a single null measurement, both the mean counts and its variance are equal to 1. The mean and variance of rate can be easily calculated from Eqs. (21).

Subsequently, we discuss CL and significance for $n = 1$. The exponential curve based on Eq. (19) is plotted in Fig. 1. Mean counts are indicated. The CL is the area under the curve to the left of upper limit $U_\theta$, while the significance $\alpha$ is the area to the right of $U_\theta$, which can be calculated from Eq. (22a). The confidence level establishes the upper limit such that, in a random experiment, the true value would be below it, and a significance that it would be above it. In Fig. 1, $U_\theta = 2.3$ corresponds to CL = 0.9 and $\alpha = 0.1$. These and other values are listed in Table 3.

The frequently used CL = 0.95 would require $\theta = 3.0$. The 1-count upper limit would give low confidence and poor significance. The upper limit for rate can be calculated from Eq. (22b).

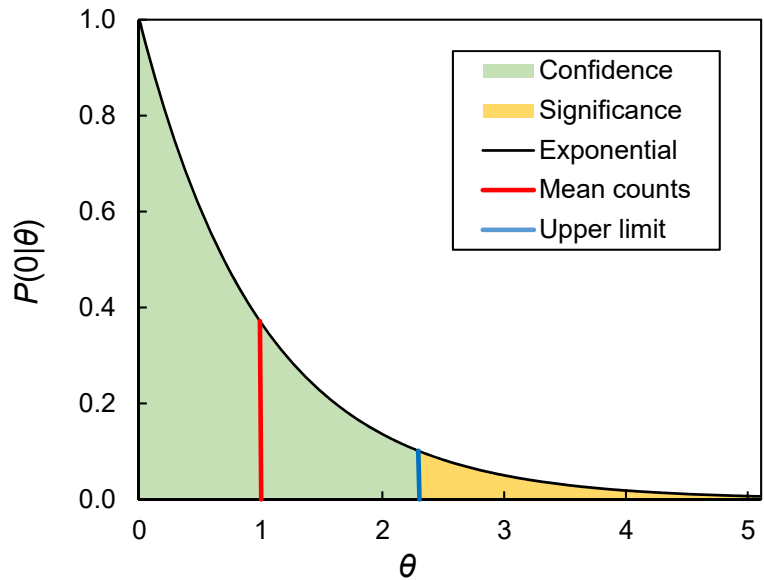

**Figure 1.** Probability of zero counts plotted as a function of $\theta$ parameter for the Poisson distribution.

**Table 3.** Upper limits for $\theta$ in counts for the *ad-hoc* probability method.

| Significance | Confidence Level | Upper limit |
|---|---|---|
| 0.01 | 0.99 | 4.6 |
| 0.05 | 0.95 | 3.0 |
| 0.10 | 0.90 | 2.3 |
| 0.37 | 0.63 | 1.0 |

Turning now to a comparison of the simple-probability method with the Bayesian analysis using uniform prior (BL), it is seen that special cases of the posterior given by Eq. (25) are: Eq. (2) when $n = 1$, Eq. (20) when $S = 0$, whereas Eq. (19) when $n = 1$ and $S = 0$. The only difference is that the condition $(x|\theta)$ is inverted to $(\theta|x)$ for the posterior. Therefore, the method called the *ad-hoc* simple-probability is not classical statistics, but Bayesian with uniform prior. Since the classical ML gives a pathological result, and the simple-probability method is essentially Bayesian, there appears to be no classical solution to the ZCD problem.

A more general Bayesian inference was provided, which included uniform prior (BL) as well as the reference priors (JJ, JR, ME), Eqs. (24,28,34,38), respectively. The properties of the priors are listed as rows in Table 2. All the reference priors are objective in a sense of the principle of indifference. If independent observers were asked to apply the underlying principles (selected from unbiased, invariant, Fisher information, and maximum entropy), as marked "yes" in the rows of Table 2, they would arrive at the same result for a given row. If no principle was selected, one would arrive at the BL prior.

The priors are plotted in Fig. 2. It is seen that the JJ and JR priors diverge at $\rho = 0$, and have long tails. The ME prior is a compromise between the BL prior and the JJ and JR at low rates, whereas it is decreasing the fastest at high rates.

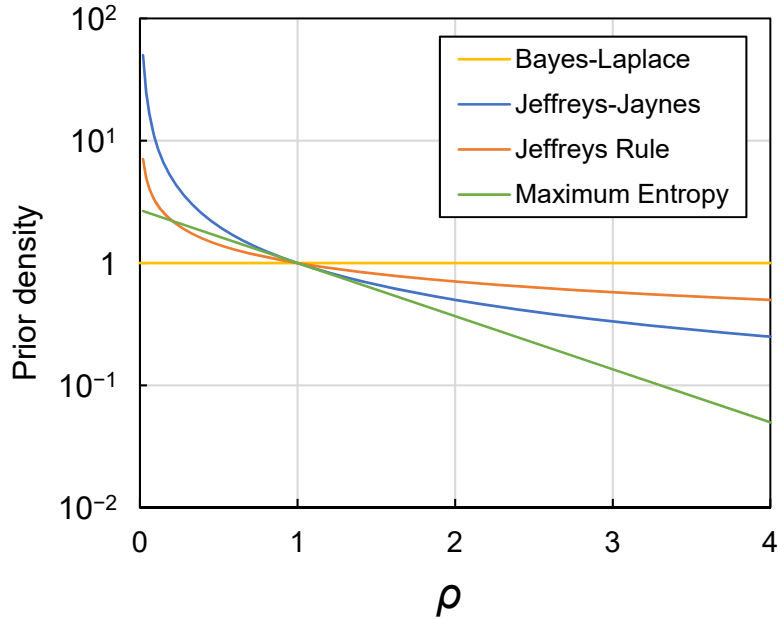

**Figure 2.** Prior density plotted as a function of $\rho$ for the Poisson distribution. The curves are normalized to pass through (1,1).

The general posterior for the rate $\rho$ is given by Eq. (43), the point estimators by Eqs. (44), and the upper limit by Eq. (45). We discuss the results for S > 0 first, which imply at least 1 count detected.

The Bayesian point estimates for the JJ prior from Eqs. (44b,c) are unbiased and identical to the classical ML (Eqs. (16) and (18c) for the mean rate and its variance, respectively). Unfortunately, there is one difficulty. The Bayesian analysis requires $A = S + a > 0$ and $B = nt + b > 0$. $B > 0$ even if $b = 0$ because $n$ and $t$ are always positive. However, for $A > 0$ one has to have at least one of $S$ or $a$ positive. For the JJ prior, we have $a = 0$ (Table 2), so we have to have $S \geq 1$ (at least 1 count detected). For the ZCD $S = 0$, so the normalization integral (Bayesian evidence) in Eq. (23) diverges. When $n = t = 1$, using Eqs. (11) and (28), the normalization integral has the form of the divergent integral $D_1$ from Eq. (64c), where variable $a$ is replaced by $\rho$.

To elucidate this further, we convert Eq. (45) to the space of counts by setting $S = 0$ and $n = t = 1$. The upper limit of the rate becomes that of the counts, $U_\theta$. One can show that, for arbitrary small $\epsilon$, Eq. (45) can be expressed as:

$$\alpha = \frac{E_1(U_\theta + \epsilon)}{-\gamma - \ln \epsilon},$$

(66)

where $E_1$ is a version of the Exponential integral and $\gamma = 0.5771 \ldots$ is Euler constant [45,64]. This equation does not have a solution for any $U_\theta$ and $\epsilon$. The JJ prior is said to be improper and has to be rejected from considerations for ZCD. We were still able to calculate Bayesian point estimates for the JJ prior. By taking the limits to $S = a = b = 0$ in Eqs. (44b, 44c), both the mean and variance are zero, which just as pathological as it was for the ML method.

In the following, we are left with three choices for the prior: BL, JR, and ME. To reveal the behavior of the posteriors, we plot them in Fig. 3 for ZCD ($S = 0$) according to Eq. (43), which was converted to counts space by setting $n = t = 1$. In this way, one can compare the behavior without time dependence. The BL posterior is the same curve as in Fig. 1.

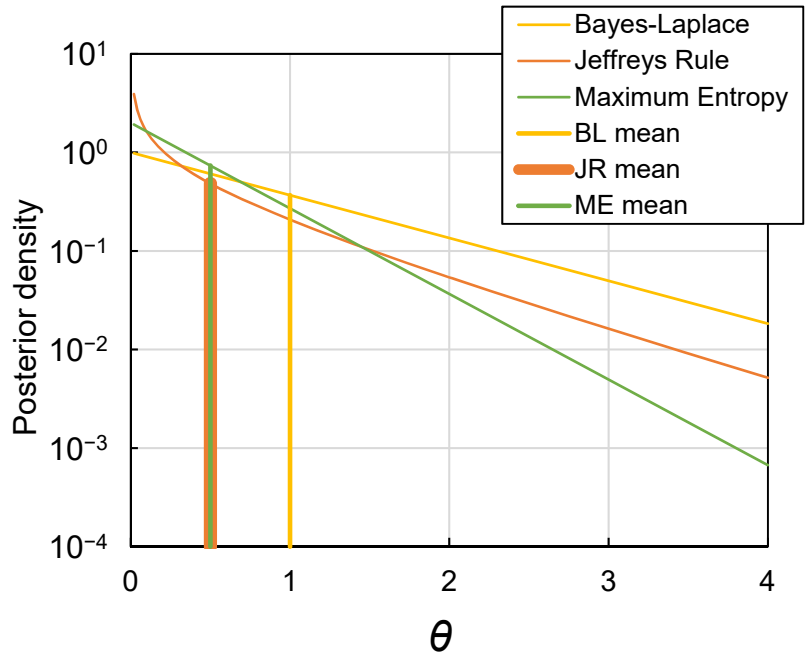

**Figure 3.** Posterior density of ZCD plotted as a function $\theta$ parameter for the Poisson likelihood. The vertical bars are Bayesian means.

The calculated Bayesian point estimates for ZCD: mean from Eq. (44b) and variance from Eq. (44c) are given in Table 4, whereas the means are also depicted in Fig. 3 as vertical bars. The mean of 1 for the BL prior appears to be some justification for taking the non-statistical 1-count upper limit described above. The mean for both JR and ME priors is equal to 1/2, whereas the variance for ME prior is the lowest at 1/4.

**Table 4.** Results of the Bayesian point estimates, bias, and risk from the Poisson likelihood for ZCD ($S = 0$) and $n = t = 1$.

| Prior | Bayesian mean | | | Bayesian variance | | |
|-------|------|------|------|----------|------|------|
| | Mean | Bias | Risk | Variance | Bias | Risk |
| BL | 1 | 1 | 1 | 1 | 1 | 1 |
| JR | 1/2 | 1/2 | 1/4 | 1/2 | 1/2 | 1/4 |
| ME | 1/2 | 1/2 | 1/4 | 1/4 | 1/4 | 1/16 |

Then, which of the three results in Table 4 should we select? This can be done based on the basis of bias and risk. The bias and risk of the Bayesian mean are calculated from Eqs. (49) and (51), respectively, whereas the bias and risk of the Bayesian variance are calculated from Eqs. (53) and (54), respectively. The results are given in Table 4. Not only the three remaining solutions for ZCD are Bayesian, but they are all biased by the definition of classical statistics. Nevertheless, Jaynes [24] found evidence of pathology in the unbiased estimates, as described above. The bias and risk are always lower for the JR and ME priors than for the BL prior. Therefore, the BL prior can be rejected. It follows from Table 4 that the JR and ME priors have the same biases and risks for the mean. To distinguish between them, one observes that the ME prior has lower bias and risk for the variance, than the JR prior. Therefore, the ME prior is the most admissible.

In order to further distinguish between the JR and ME priors, one can extend the principle of indifference by elucidating the underlying principles from Table 2 as follows. The JR prior has one "yes" in Table 2, whereas the ME prior has two. Therefore, the ME prior contains more information, so any objective observer would select it over the JR prior. Consequently, the ME prior provided the most acceptable solution to the ZCD problem.

Bayesian upper limits in units of counts are calculated by solving Eq. (45) for $S = 0$ and $n = t = 1$. They are plotted in Fig. 4. Specific values for credibility levels of 0.90, 0.95, and 0.99 are listed in Table 5. It is seen from Fig. 4 and Table 5 that the upper limits increase as the credibility level increases. In addition, for high credibility levels, the upper limits decrease as we move from the BL to JR and to ME prior.

**Table 5.** Bayesian upper limits in counts.

| Bayesian credibility level | Prior | | |
|---|---|---|---|
| | BL | JR | ME |
| 0.90 | 2.3 | 1.4 | 1.2 |
| 0.95 | 3.0 | 1.9 | 1.5 |
| 0.99 | 4.6 | 3.3 | 2.3 |

We also described two recognized statistical distributions of potential application to ZCD: zero-inflated Poisson, $P_z(x|\theta, \psi)$, and Negative binomial, $NB(x|\theta, a)$. The z-Poisson has enhanced probability for zero class. The NB is an inverse of the binomial distribution. They both can be derived from the Poisson distribution. The probability mass functions (pmf) for the three distributions are depicted in Fig. 5, for a common mean equal to 4 and dispersion coefficients equal to 1.5. This overdispersion is visible in Fig. 5.

The z-Poisson and NB are both 2-parameter distributions. In order to estimate both parameters, one needs more than one measurement and non-zero data point(s). However, this is not available for ZCD as defined in this paper, since all results are null. Therefore, there is no handle to apply these distributions to ZCD. The question is: what is the implication of using the Poisson distribution, while the true hidden distribution might be that of either z-Poisson on NB, which is not known to the observer? This conundrum can be resolved by using the rules of Bayesian statistics, which allow elimination of unknown (also called nuisance) parameters by integrating them out. This process is called marginalization. In Section 4, we assumed ME priors and marginalized the z-Poisson and NB posteriors over the $\psi$ and $a$ parameters, using Eqs. (58) and (65), respectively. The result in both cases was the Poisson posterior, Eq. (43). The interpretation is that, even if the actual distribution was other than the Poisson, we can use the Poisson distribution for ZCD as long as the actual distribution derives from it. This feature in not available in classical statistics and is a significant advantage of the Bayesian statistics.

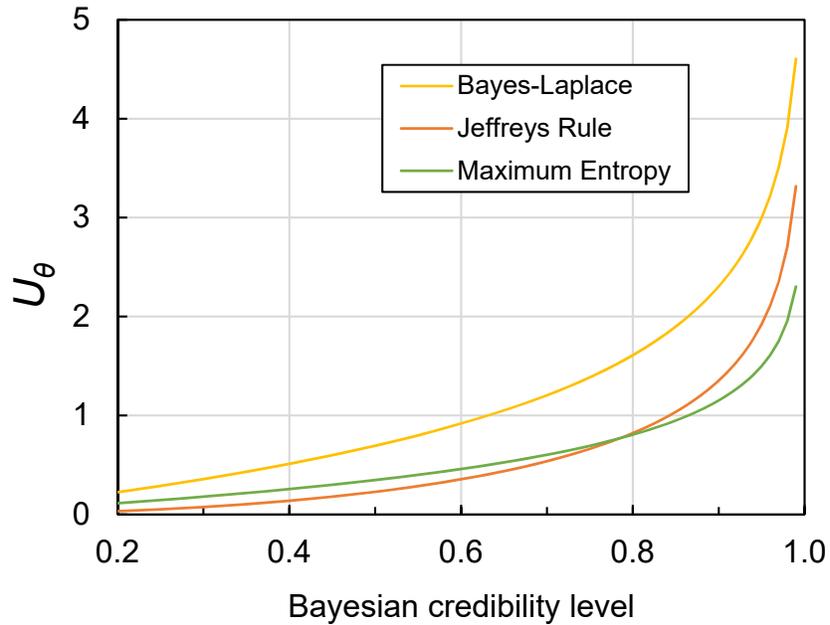

**Figure 4.** Bayesian upper limits, $U_\theta$, plotted as a function of Bayesian credibility level for ZCD.

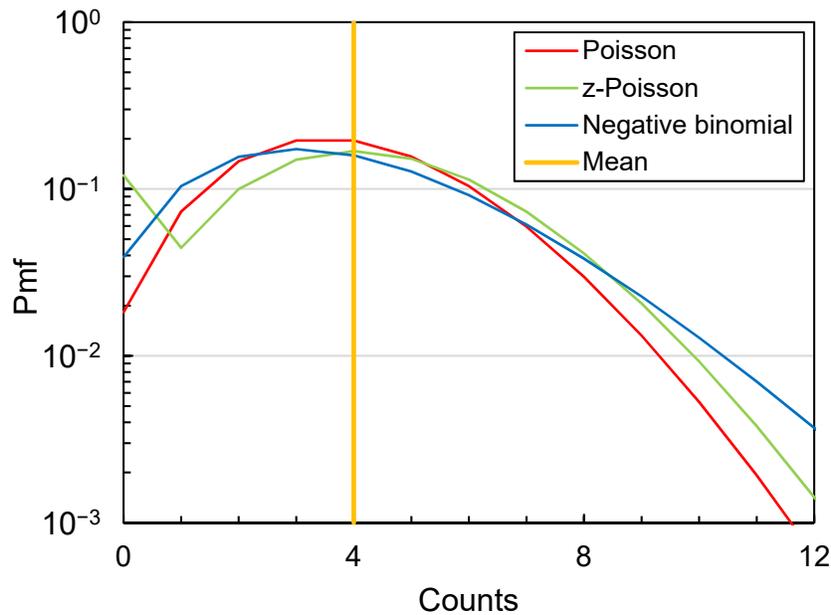

**Figure 5.** Probability mass functions for Poisson, z-Poisson, and Negative-binomial distributions for mean of 4 and $\delta_x = 1.5$.

## 6. SUMMARY AND CONCLUSIONS

In this work, we developed a consistent statistical analysis of Zero-Count Detector. The ZCD is a detector sensitive to stimulus, however, the results of either a single or multiple measurement are null under condition that the background is also null, not measured, or rejected by event. The ZCD has applications across many fields where rare events are studied. The goal of such studies is to set an upper limit on the rate of physical process, contingent upon observing null result.

The approach in this paper was based on the Poisson likelihood. The properties of the Poisson distribution were reviewed, such as count parameter, rate, statistical moments (mean and variance), and dispersion coefficient. Example was given in radiation counting from a long-lived radionuclide, which also included detection efficiency. Deviations from the Poisson distribution were described in terms of under- and over-dispersed statistics.

After briefly discussing the non-statistical 1-count upper limit, we focused on a classical statistical analysis and the Bayesian analysis of the Poisson likelihood.

Within the classical statistics, we discussed properties such as sufficient statistics and bias. Using the method of Maximum Likelihood, we showed that the ML estimates for the mean and variance are unbiased. When applied to ZCD, these estimates were zero. Whereas zero mean would be acceptable, zero variance does not allow any estimation of either uncertainty or an upper limit. This result is pathological. Another seemingly classical statistics concept was investigated based on an *ad-hoc* probability of zero-class. It was later showed that this was essentially a Bayesian statistics with uniform prior. Therefore, there appears to be no solution to ZCD within the boundaries of classical statistics.

We described essential properties and rules of the Bayesian statistics, such as prior, posterior, evidence, marginalization, and the principle of indifference. Since no information other than zero counts and duration of measurement are available for ZCD, one has to resort to theoretical priors in the Bayesian approach. These are either uniform (Bayes-Laplace) prior or the reference priors. Among the latter, we investigated the Jeffreys-Jaynes prior based on the transformation invariance, the Jeffreys Rule based on Fisher information, and the Maximum Entropy prior based on the very principle. Using these priors, we derived posterior in a form of Gamma distribution, as well as calculated the Bayesian moments and upper limits for both the count parameter and the rate.

While the JJ-prior-derived posterior is the only one among Bayesian that leads to unbiased estimates, it is only valid when at least one count was detected. For zero counts, it leads to Bayesian evidence being a divergent integral so that the posterior cannot be normalized to probability. Therefore, the JJ prior is improper and cannot be used for ZCD.

Among the remaining BL, JR, and ME priors, the BL prior leads to the highest bias and risk for both the mean and variance. Therefore, it can be rejected based on admissibility. The ME prior leads to the same bias and risk for the mean as the JR prior. However, it has lower bias and risk for the variance. Thus, the ME prior is the most admissible among the priors studied. In addition, the ME prior is based on the most information and, using the principle of indifference, it is the most acceptable.

While the Poisson distribution was showed to be an efficient likelihood for the ZCD studies, it is by no means the only one possible. We investigated two available distributions: zero-inflated

Poisson and Negative binomial, which may better describe the fluctuating physical process. Both distributions derive from the Poisson, have two parameters each, and are overdispersed. Whether using classical or Bayesian statistics, inference from such distribution requires measurement results more than zero, which is not available for ZCD. Nevertheless, the Bayesian rule allows marginalization over one of the two parameters. It was then showed that the posterior based on either the z-Poisson or the NB likelihood reduces to the Poisson posterior upon marginalization.

**Acknowledgements**

This work was partially supported by the US FDA of the US Department of Health and Human Services financial assistance award U19FD007089. The late X. Li is acknowledged for his role in designing of this project.

**References**

[1] Loredo T.J., Lamb D.Q. (2002) Bayesian analysis of neutrinos observed from supernova SN 1987A. Phys. Rev. D 65, 063002-1-39. https://doi.org/10.1103/PhysRevD.65.063002

[2] Justus A. (2019) Multiple facets of the Poisson distribution applicable to health physics measurements. Health Phys. 117, 36-57. https://doi.org/10.1097/HP.0000000000001013

[3] Barry S.C., Welsh A.H. (2002) Generalized additive modelling and zero inflated count data. Ecolog. Model. 157, 179-188. https://doi.org/10.1016/S0304-3800(02)00194-1

[4] Chin H.C., Quddus M.A. (2003) Modeling count data with excess zeroes an empirical application to traffic accidents. Sociol. Meth. Res. 32, 90-116. https://doi.org/10.1177/0049124103253459

[5] Özmen I., Famoye F. (2007) Count regression models with an application to zoological data containing structural zeros. J. Data Sci. 5, 491-502. https://doi.org/10.6339/JDS.2007.05(4).385

[6] Francois M., Peter C., Gordon F. (2012) Dealing with excess of zeros in the statistical analysis of magnetic resonance imaging lesion count in multiple sclerosis. Pharmaceut. Statist. 11, 417-424. https://doi.org/10.1002/pst.1529

[7] Loeys T., Moerkerke B., De Smet O., Buysse A. (2012) The analysis of zero-inflated count data: beyond zero-inflated Poisson regression. British J. Math. Statist. Psych. 65, 163-180. https://doi.org/10.1111/j.2044-8317.2011.02031.x

[8] Cagigal M.P., Canales V.F. (2001) Exoplanet detection using a nulling interferometer. Optics Expr. 9, 36-41. https://doi.org/10.1364/OE.9.000036

[9] Semkow T.M. (2006a) Exponential decay law and nuclear statistics. In Applied Modeling and Computations in Nuclear Science, A.C.S. Symposium Series 945, Ed. by Semkow T.M., Pommé S., Jerome S., Strom D.J. Washington, DC, pp. 42-56. https://doi.org/10.1021/bk-2007-0945.fw001

[10] Johnson N.L., Kemp A.W., Kotz S. (2005) Univariate Discrete Distributions, 3rd ed. Wiley-Interscience, Hoboken. https://doi.org/10.1002/0471715816

[11] Beach S.E., Semkow T.M., Remling D.J., Bradt C.J. (2017) Demonstration of fundamental statistics by studying timing of electronics signals in a physics-based laboratory. Am. J. Phys. 85, 515-521. https://doi.org/10.1119/1.4982165

[12] Semkow T.M. (2002) Experimental verification of overdispersion in radioassay data. Health Phys. 83, 485-496. https://doi.org/10.1097/00004032-200210000-00006

[13] Pommé S. (1999) How pileup rejection affects the precision of loss-free counting. Nucl. Instr. Meth. Phys. Res. A 432, 456-470. https://doi.org/10.1016/S0168-9002(99)00475-1

[14] Jenkins P.H., Burkhart J.F., Kershner C.J. (2006) Corrections for overdispersion due to correlated counts in radon measurements using grab scintillation cells, activated charcoal devices and liquid scintillation charcoal devices. In Applied Modeling and Computations in Nuclear Science, A.C.S. Symposium Series 945, Ed. by Semkow T.M., Pommé S., Jerome S., Strom D.J. Washington, DC, pp. 249-268. https://doi.org/10.1021/bk-2007-0945.fw001

[15] Mekjian A.Z. (2001) Model for studying branching processes, multiplicity distributions, and non-Poissonian fluctuations in heavy-ion collisions. Phys. Rev. Lett. 86, 220-223. https://doi.org/10.1103/PhysRevLett.86.220

[16] Salma I., Zemplén-Papp É. (1992) Experimental investigation of statistical models describing distribution of counts. Nucl. Instr. Meth. Phys. Res. A 312, 591-597. https://doi.org/10.1016/0168-9002(92)90209-M

[17] Lee S.H., Jae M., Gardner R.P. (2007) Non-Poisson counting statistics of a hybrid G-M counter dead time model. Nucl. Instr. Meth. Phys. Res. B 263, 46-49. https://doi.org/10.1016/j.nimb.2007.04.041

[18] Anderson J.L. (1972) Non-Poisson distributions observed during counting of certain carbon-14-labeled organic (sub)monolayers. J. Phys. Chem. 76, 3603-3612. https://doi.org/10.1021/j100668a018

[19] Funk A.C., Beck M. (1997) Sub-Poissonian photocurrent statistics: theory and undergraduate experiment. Am. J. Phys. 65, 492-500. https://doi.org/10.1119/1.18577

[20] Blain E., Nishikawa K., Faye S.A, Roselan A. Burn A.G., Torres M.A., Semkow T.M. (2023) Detection capability of $^{89}$Sr and $^{90}$Sr using liquid scintillation counting. Health Phys. 125, 123-136. http://dx.doi.org/10.1097/HP.0000000000001698

[21] Semkow T.M., Li X., Chu L.T. (2019) Overview of signal detection theory and detection limits. In Detection Limits in Air Quality and Environmental Measurements, Ed. by Brisson M.J, American Society for Testing and Materials International, West Conshohocken, pp. 60-76. http://dx.doi.org/10.1520/STP161820180061

[22] Semkow T.M. (2006b) Bayesian inference from the binomial and Poisson processes for multiple sampling. In Applied Modeling and Computations in Nuclear Science, A.C.S. Symposium Series 945, Ed. by Semkow T.M., Pommé S., Jerome S., Strom D.J. Washington, DC, pp. 335-356. https://doi.org/10.1021/bk-2007-0945.fw001

[23] Stuart A., Ord K., Arnold S. (1999) Kendall's Advanced Theory of Statistics, Vol. 2A, Classical Inference and the Linear Model, 6th. ed. Arnold, London.

[24] Jaynes E.T. (2003) Probability Theory, the Logic of Science, Ed. by Bretthorst G.L. Cambridge U.P., Cambridge. https://doi.org/10.1017/CBO9780511790423

[25] Johnson N.L. (1951) Estimators of the probability of the zero class in Poisson and certain related populations. Ann. Math. Statist. 22, 94-101. https://doi.org/10.1214/aoms/1177729696

[26] Leo W.R. (1994) Techniques for Nuclear and Particle Physics Experiments, 2nd ed. Springer-Verlag, Berlin. https://doi.org/10.1007/978-3-642-57920-2

[27] Loveland W.D., Morrisey D.J., Seaborg G.T. (2017) Modern Nuclear Chemistry, 2nd ed. Wiley, Hoboken. https://doi.org/10.1002/9781119348450

[28] Berger J.O. (1985) Statistical Decision Theory and Bayesian Analysis, 2nd ed. Springer-Verlag, New York. https://doi.org/10.1007/978-1-4757-4286-2

[29] Box G.E.P., Tiao G.C. (1992) Bayesian Inference in Statistical Analysis. J. Wiley & Sons, New York. https://doi.org/10.1002/9781118033197

[30] O'Hagan A., Forster J. (2004) Kendall's Advanced Theory of Statistics, Vol. 2B, Bayesian Inference, 2nd ed. Arnold, London.

[31] Bernardo J.M., Smith A.F.M. (2004) Bayesian Theory. J. Wiley & Sons, Chichester.

[32] Gelman A., Carlin J.B., Stern H.S., Dunson D.B., Vehtari A., Rubin D.B. (2014) Bayesian Data Analysis, 3rd ed. CRC Press, Boca Raton. https://doi.org/10.1201/b16018

[33] Bayes T. (1763) An essay towards solving a problem in the doctrine of chances. Phil. Trans. Roy. Soc. (London) 53, 370-418. Reprinted with an introduction by Barnard G.A. (1958) Biometrika 45, 293-315. https://doi.org/10.2307/2333180

[34] Laplace P.S. Mémoire sur la probabilité des causes par les évènemens (1774) Mémoires de mathématique et de physique presentés à l'Académie royale des sciences, par divers savans, & lús dans ses assemblées 6, 621-656. Reprinted with an introduction and English translation by Stigler S.M. (1986) Stat. Sci. 1, 359-378.

[35] Jeffreys H. (1961) Theory of Probability, 3rd ed. Clarendon Press, Oxford.

[36] Robert C.P., Chopin N., Rousseau J. (2009) Harold Jeffreys's Theory of Probability revisited. Statist. Sci. 24, 141-172. https://doi.org/10.1214/09-STS284

[37] Gorroochurn P. (2016) Classic Topics on the History of Modern Mathematical Statistics: From Laplace to More Recent Times. J. Wiley & Sons, Hoboken. https://doi.org/10.1002/9781119127963

[38] Jaynes E.T. (1983) Papers on Probability, Statistics and Statistical Physics, Ed. by Rosenkrantz R.D. D. Reidel Publishing Company, Dordrecht.

[39] Prosper H.B. (1988) Small-signal analysis in high-energy physics: a Bayesian approach. Phys. Rev. D. 37, 1153-1160. https://doi.org/10.1103/PhysRevD.37.1153

[40] Toussaint U.v. (2011) Bayesian inference in physics. Rev. Mod. Phys. 83, 943-999. https://doi.org/10.1103/RevModPhys.83.943

[41] Little R.J.A. (1982) The statistical analysis of low-level radioactivity in the presence of background counts. Health Phys. 43, 693-703.
https://doi.org/10.1097/00004032-198211000-00007

[42] Miller G., Martz H.F., Little T.T., Guilmette R. (2002) Using exact Poisson likelihood function in Bayesian interpretation of counting measurements. Health Phys. 83, 512-518.
https://doi.org/10.1097/00004032-200210000-00009

[43] In Ref. [38], Where do we stand on maximum entropy? (1978) pp. 210-314.

[44] Friedlander G., Kennedy J.W., Macias E.S., Miller J.M. (1981) Nuclear and Radiochemistry, 2nd ed. J. Wiley & Sons, New York.

[45] Gradshteyn I.S., Ryzhik I.M. (2015) Table of Integrals, Series and Products, 8th ed., Ed. by Zwillinger D., Moll V. Academic Press, San Diego.

[46] Jeffreys H. (1957) Scientific Inference, 2nd ed. Cambridge University Press, London.

[47] Jeffreys H. (1932) On the theory of errors and least squares. Proc. Royal Soc. London A 138, 48-55. https://doi.org/10.1098/rspa.1932.0170

[48] Gull S.F. (1988) Bayesian inductive inference and maximum entropy. In Maximum-Entropy and Bayesian Methods in Science and Engineering, Vol. 1: Foundations, Ed. by Erickson G.J., Smith C.R., Kluver A.P., Dordrecht, pp. 553-74. https://doi.org/10.1007/978-94-009-3049-0_4

[49] In Ref. [38], Prior probabilities (1968) pp. 114-130.

[50] Haldane J.B.S. (1948) The precision of observed values of small frequencies. Biometrika 35, 297-300. https://doi.org/10.1093/biomet/35.3-4.297

[51] Jeffreys H. (1946) An invariant form for the prior probability in estimation problems. Proc. Roy. Soc. London A 186, 453-461. https://doi.org/10.1098/rspa.1946.0056

[52] Perks W. (1947) Some observations on inverse probability including a new indifference rule. J. Inst. Actuaries 73-285-334. https://doi.org/10.1017/S0020268100012270

[53] Bernardo J.M. (1979) Reference posterior distributions for Bayesian inference. J. Roy. Stat. Soc. B 41, 113-147. https://doi.org/10.1111/j.2517-6161.1979.tb01066.x

[54] In Ref. [38], Information theory and statistical mechanics (1963) pp. 39-76.

[55] In Ref. [38], Confidence intervals vs Bayesian intervals (1976) pp. 139-209.

[56] Lindley D.V. (1956) On a measure of the information provided by the experiment. Ann. Math. Statist. 27, 986-1005. https://doi.org/10.1214/aoms/1177728069

[57] Fisher R.A. (1933) The concepts of inverse probability and fiducial probability referring to unknown parameters. Proc. Royal Soc. London A 139, 343-348.
https://doi.org/10.1098/rspa.1933.0021

[58] Kass R.E., Wasserman L. (1996) The selection of prior distributions by formal rules. J. Am. Statist. Assoc. 91, 1343-1370. https://doi.org/10.1080/01621459.1996.10477003

[59] Hartigan J. (1964) Invariant prior distributions. Ann. Math. Statist. 35, 836-845.

https://doi.org/10.1214/aoms/1177703583

[60] Akaike H. (1978) A new look at the Bayes procedure. Biometrika 65, 53-59. https://doi.org/10.1093/biomet/65.1.53

[61] Kerman J. (2011) Neutral noninformative and informative conjugate beta and gamma prior distributions. Electron. J. Statist. 5, 1450-1470. https://doi.org/10.1214/11-EJS648

[62] Irony T.Z. Bayesian estimation for discrete distributions (1992) J. Appl. Statist. 19, 533-549. https://doi.org/10.1080/02664769200000049

[63] Johnson N.L., Kotz S., Balakrishnan N. (1994) Continuous Univariate Distributions, 2nd ed., Vol. 1. J. Wiley & Sons, New York.

[64] Gautschi W., Cahil W.C. (1972) Exponential integral and related functions. In Handbook of Mathematical Functions, Ed. by M. Abramowitz, I.A. Stegun, Dover Publications, New York, pp. 227-254.

**Symbol glossary**

*Abbreviations*

| | |
|---|---|
| BL | Bayes-Laplace |
| CL | confidence level, credibility level |
| JJ | Jeffreys-Jaynes |
| JR | Jeffreys rule |
| ME | maximum entropy |
| ML | maximum likelihood |
| NB | Negative binomial |
| pdf | probability density function |
| pmf | probability mass function |
| UMV | unbiased minimum variance |
| Var | variance |
| ZCD | zero-count detector |
| z-Poisson | zero-inflated Poisson |

*Variables*

| | |
|---|---|
| $a, b$ | Gamma prior parameters |
| $(a)_r$ | ascending factorial |
| $A, B$ | Gamma posterior parameters |
| $c$ | constant |
| $d$ | exponent |
| $D_y$ | divergent integral of the $y$-th order |
| $E[\ ]$ | expectation value |
| $E_1$ | Exponential integral |
| $F$ | functional |
| $_2F_1$ | Gaussian hypergeometric function |
| $g$ | function |
| $h$ | function, invariant measure |
| $i, j, k$ | indices |
| $I_{x+k}, K_{x+k}$ | symbols for integrals |
| $J$ | Fisher information |
| $L$ | likelihood function |
| $m$ | number of groups, exponent |
| $n$ | number of measurements |
| $N$ | number of events, radioactive atoms |
| $p$ | probability |
| $P$ | probability, probability density, prior, posterior |
| $P_j$ | abbreviation for $P(j|\theta)$ |
| $P_z$ | z-Poisson |
| $q$ | constant |
| $S$ | sum of counts |
| $t, t'$ | measurement time, rescaled time |
| $U, U_\rho, U_\theta$ | upper limit, of rate, of counts |
| $v$ | excess variation coefficient |
| $V_B$ | Bayesian variance |
| $W$ | entropy |
| $x, x_i$ | Poisson random variable (counts) |
| $\boldsymbol{x}$ | count-data vector |
| $\bar{x}$ | mean counts |
| $y$ | index |
| | |
| $\alpha$ | significance |
| $\beta$ | parameter |
| $\gamma$ | Incomplete Gamma function, Euler constant |
| $\Gamma$ | Gamma function |
| $\delta_x$ | dispersion coefficient of counts |
| $\varepsilon$ | detection efficiency |
| $\epsilon$ | arbitrary small constant |
| $\theta, \hat{\theta}, \theta_B$ | Poisson parameter, ML estimate, Bayesian estimate |
| $\kappa, \nu$ | Lagrange multipliers |
| $\lambda$ | radioactive decay constant |
| $\mu, \hat{\mu}$ | mean, ML estimate |
| $\boldsymbol{\mu}$ | mean-data vector |
| $\mu_1'$ | 1-st uncorrected moment (mean) |
| $\mu_r'$ | $r$-th uncorrected moment |
| $\mu_2, \hat{\mu}_2$ | 2-nd corrected moment (variance), ML estimate |
| $\rho, \rho', \hat{\rho}, \rho_0$ | Poisson rate, rescaled rate, ML estimate, mean rate |
| $\sigma$ | standard deviation |
| $\varphi$ | parameter |
| $\psi$ | z-Poisson parameter |